\documentclass[aps,prb,twocolumn,superscriptaddress,showpacs,letter]{revtex4}
\usepackage{epsfig,graphicx,times}
\newcommand{\mb}[1]{\mathbf{#1}}
\newcommand{\mr}[1]{\mathrm{#1}}
\begin{document}
%\preprint{}
\title{Electron-Electron Interactions in Isolated and Realistic Quantum Dots:\\
 A Density Functional Theory Study}
\author{Hong Jiang}
\email[]{hong.jiang@duke.edu}
\affiliation{Department of Chemistry, Duke University, Durham, North Carolina 27708-0354}
\affiliation{Department of Physics, Duke University, Durham, North Carolina 27708-0305}
\affiliation{College of Chemistry and Molecular Engineering, Peking University, Beijing, China 100871}
\author{Denis Ullmo}
\email[]{ullmo@phy.duke.edu}
\affiliation{Department of Physics, Duke University, Durham, North Carolina 27708-0305}

\author{Weitao Yang}
\email[]{weitao.yang@duke.edu}
\thanks{Corresponding authors}
\affiliation{Department of Chemistry, Duke University, Durham, North Carolina 27708-0354}

\author{Harold U. Baranger}
\email[]{baranger@phy.duke.edu}
\thanks{Corresponding authors}
\affiliation{Department of Physics, Duke University, Durham, North Carolina 27708-0305}

\date{\today}

\begin{abstract}
We use Kohn-Sham spin-density-functional theory to study the statistics of ground-state spin and the spacing between conductance peaks in the Coulomb blockade regime for both 2D isolated and realistic quantum dots. We make a systematic investigation of the effects of electron-electron interaction strength and electron number on both the peak spacing and spin distributions. A direct comparison between the distributions from isolated and realistic dots shows that, despite the difference in the boundary conditions and confining potential, the statistical properties are qualitatively the same.
Strong even/odd pairing in the peak spacing distribution is observed only in the weak e-e interaction regime and vanishes for moderate interactions. The probability of high spin ground states increases for stronger e-e interaction and seems to saturate around $r_s \!\sim\! 4$. The saturated value is larger than previous theoretical predictions. Both spin and conductance peak spacing distributions show substantial variation as the electron number increases, not saturating until $N \!\sim\! 150$. To interpret our numerical results, we analyze the spin distribution in the even $N$ case using a simple two-level model.  

\end{abstract}
%\keywords{}
\maketitle

\section{Introduction}

The interplay of quantum mechanical interference and electron-electron interaction in semiconductor quantum dots (QDs) has attracted  a lot of experimental and theoretical interest in recent years.\cite{Kouwenhoven97,Alhassid00RMP,Aleiner02,Reimann02RMP} Experimentally the ground state properties of a QD can be probed by Coulomb blockade (CB) measurements at near-zero temperature, in which the conductance through the dot is blocked except at particular gate voltages, causing the conductance as a function of gate voltage to be a series of sharp peaks. In particular, the spacing between neighboring peaks is related to the second difference of the ground state energy with respect to electron number $N$, $\Delta_2 E(N)\!\equiv\!E_\mr{gs}(N+1)\!+\!E_\mr{gs}(N-1)\!-\!2 E_\mr{gs}(N)$, which is often called the addition energy. The ground state spin of the dot can be inferred from the motion of the peak position in a magnetic field.\cite{Folk01,Lindemann02} Except in small dots with high geometrical symmetry,\cite{Tarucha96} whose ground state properties show shell structure following Hund's rules, most experimental QDs are asymmetric due to either the irregular shape of the confining gate or impurity disorder, and the number of electrons is usually several hundred.\cite{Sivan96,Patel98,Simmel97,Luscher01} The properties of QDs show apparently random but reproducible fluctuations as some external parameters such as the chemical potential, the shape of the confining potential, external magnetic field and so on, vary. In this case, only statistical properties are of physical significance.

While the statistical behavior of peak height fluctuations can be well described by a non-interacting random matrix theory (RMT) model,\cite{Alhassid00RMP} a similar model, the constant interaction (CI) plus RMT, fails dramatically in describing the peak spacing distribution. The importance of e-e interaction was well recognized, but different theoretical approaches differ quantitatively or even qualitatively in many aspects of e-e interactions effects. The numerical studies for small dots\cite{Sivan96,Prus96,Berkovits98,Cohen99,Walker99,Levit99,Ahn99,Bonci99,Hirose02} yield a Gaussian-like peak spacing distribution without any even/odd pairing effect, which agrees with experimental results on much larger dots. On the other hand, the statistical approaches for large dots\cite{Ullmo01b,Usaj01,Usaj02} predict a much more pronounced even/odd effect at zero temperature that is absent in experimental results.\cite{remark-Usaj}

Besides conductance peak spacings, another quantity that has attracted a lot of interest is the ground state spin in the CB regime. The occurrence of non-minimal ground state spin has been suggested as an explanation for the absence of a bi-modal distribution in conductance peak spacings,\cite{Prus96} and for the kinks in the parametric motion of the peaks.\cite{Baranger00} A recent experimental study on the change in CB peak position as a function of magnetic filed by Folk, et al.\cite{Folk01} found that three out of five even-$ N $ states had $ S \!=\! 1 $. As in the case of peak spacing distributions, previous theoretical studies on ground state spin distributions in QDs gave very different results: Simulations on small tight-binding lattice models with random site energies  predict a dominant fraction of high spin ground states,\cite{Kurland00} but results from KS-SDFT studies of small disordered parabolic quantum dots\cite{Hirose02} showed that the probability of high-spin ground states never exceeds $50\%$ even in the strong interaction regime.

Despite intensive efforts to understand the role of e-e interaction in QD systems, there are still many questions that require further investigation. In this paper, we focus on the following issues: (1) The first one is concerned with the evolution of statistical properties of QDs as the interaction strength increases. The e-e interaction strength is usually characterized by the Wigner-Seitz radius $r_s$, which is defined in 2D as $r_s\equiv 1/\sqrt{\pi n}a_0 $, where $n$ is the electron density and $a_0$ is Bohr radius.\cite{remark-rs} Previous theoretical studies gave different answers to this question, indicating the necessity of further theoretical exploration. In the meantime, this issue has attracted the experimental interest, and some relevant experiments are underway. (2) The second issue to be explored is the effect of confining potential. Previous theoretical studies were based on abstract statistical assumptions, chaotic model external potentials, or Gaussian impurities. None of them have considered the realistic confining potential imposed in actual experiments.  Though it is usually assumed that the statistical properties of QDs are of universal significance, the validity of that is still to be verified. (3) The third issue that is worthy of further elaboration is how statistical properties of QDs evolve  as the size of a QD system, mainly the electron number, increases, which is important not only in solving the puzzling discrepancy between numerical small-dot calculations and semi-analytic large-dot statistical analysis, but also in understanding the fundamental question in mesoscopic physics of how macroscopic properties emerge from microscopic physics via the mesoscopic regime.

In our previous study,\cite{Jiang03,Jiang03b} we reported a Kohn-Sham spin density functional theory (KS-SDFT) study of spin and conductance peak spacing distributions in 2D isolated quantum dots in a classically chaotic quartic oscillator potential with electron number $N$ up to 200 at $r_s \!\sim\! 1.5$. We found that for large electron number and asymmetric external potential, the peak spacing distribution is mainly Gaussian-like and there is no observable even/odd effects. We also found enhanced probability of high spin ground state implying stronger interaction effects than expected from RPA approaches. In this paper, we extend the previous study in two directions: (1) Similar calculations are done for weaker ($r_s \!\sim\! 0.2$) interactions to investigate how statistical properties of quantum dots evolve as the interaction strength varies. (2) We extended previous studies in isolated QDs, in which electrons are confined by an analytic chaotic potential without explicit consideration of the coupling with external gate, to QD systems in which the confining potential is calculated by solving the electrostatic Poisson equation with realistic boundary conditions. In the latter case, with more controllable parameters available, we are able to study statistical properties of QDs as a function of both e-e interaction strength ($r_s$) and electron number ($N$) in a more systematic way.

The outline of the paper is as follows: In the next section, the KS-SDFT method and the numerical techniques used in our study are concisely described. In Section III, we report the statistical properties of isolated quantum dots. In Section IV, the results for realistic quantum dots are presented. Section V concludes the paper with discussion about the significance of these results.

\section{Method}

\subsection{Kohn-Sham spin density functional theory}

In KS-SDFT, the ground state energy of an interacting system with
electron number $N$ and total spin $S$ in the local external
potential \( V_{\mathrm{ext}}(\mathbf{r}) \) is written as a
functional of spin densities \( n^{\sigma } \) with \( \sigma
=\alpha, \beta \) denoting spin-up and spin-down, respectively,
\begin{eqnarray}
E\left[ n^{\alpha },n^{\beta }\right] =T_{s}\left[ n^{\alpha
},n^{\beta }\right]
+\int n(\mb{r})V_\mathrm{ext}(\mb{r})d\mb{r} &  & \nonumber \\
+\frac{1}{2}\int \frac{n(\mb{r})n(\mb{r}')}{\left|
\mb{r}-\mb{r}'\right| }d\mb{r}d\mb{r}' +E_{\mathrm{xc}}\left[
n^{\alpha },n^{\beta }\right]  &. \label{eq: E0}
\end{eqnarray}
(Effective atomic units are used through the paper: for
GaAs QDs with the dielectric constant $\varepsilon=12.9$ and effective electron mass $m^*=0.067 m_e$, the values are 10.08 meV for energy and 10.95 nm for length.)
 \( T_{s}\left[ n^{\alpha },n^{\beta }\right] \) is the kinetic energy of
the KS non-interacting reference system which has the same ground
state spin density as the interacting one, and \(
E_\mathrm{xc}\left[ n^{\alpha },n^{\beta }\right]  \) is the
exchange-correlation energy functional.  The spin densities
$n^{\sigma }$ satisfy the constraint $\int
n^{\sigma}(\mb{r})d\mb{r}=N^{\sigma}$ with $ N^\alpha=(N+2S)/2$
and $N^\beta = (N-2S)/2$.

Assuming that the ground state of the non-interacting reference
system is non-degenerate, the non-interacting kinetic energy is
given by $T_s\left[ n^{\alpha },n^{\beta }\right]=\sum _{i,\sigma
} \left\langle \psi _{i}^{\sigma }\left| -\frac{1}{2}\nabla
^{2}\right| \psi _{i}^{\sigma }\right\rangle $, and the ground
state spin density is uniquely expressed as
\begin{equation}
n^{\sigma }(\mb{r})=\sum _{i}^{N^\sigma}\left| \psi _{i}^{\sigma
}(\mb{r})\right| ^{2}, \quad \sigma=\alpha,\beta.
\end{equation}
Here $\psi_{i}^{\sigma}$ are the lowest single-particle orbitals
which are obtained from
\begin{equation}
\label{eq:KS}  \big\lbrace -\frac{1}{2}\nabla ^{2}+V_\mathrm{ext}(\mb{r})+V_{H}[n;\mb{r}]
+V_{\mathrm{xc}}^{\sigma }[n^{\alpha },n^{\beta };\mb{r}]\big\rbrace \psi _{i}^{\sigma
}(\mb{r})=\varepsilon _{i}^{\sigma } \psi _{i}^{\sigma }(\mb{r}),
\end{equation}
where \( V_{H}[n;\mb{r}] \) and \( V_{\mathrm{xc}}^\sigma[n^{\alpha
},n^{\beta };\mb{r}]\) are the Hartree and exchange-correlation
potentials, respectively,
\begin{eqnarray}
\label{eq:VH}
V_{H}[n;\mb{r}]& \equiv &\int \frac{n(\mb{r}')}{|\mb{r}-\mb{r}'|}d^{3}\mb{r}', \\
V_{\mathrm{xc}}^\sigma[n^{\alpha },n^{\beta }; \mb{r}] & \equiv &
\frac{\delta E_\mathrm{xc} \left[ n^{\alpha} ,n^{\beta }\right]}{
\delta n^{\sigma}(\mb{r})}.
\end{eqnarray}

We have used the local spin-density approximation
(LSDA)\cite{ParrYang89,DreizlerGross90} for the exchange-correlation
energy functional,
\begin{equation}
E_\textrm{xc}\left[ n^\alpha,n^\beta \right]\approx \int
n(\mb{r})\epsilon_\textrm{xc} \big(n(\mb{r}),\zeta(\mb{r})\big)d\mb{r},
\end{equation}
where $\zeta=(n^\alpha-n^\beta)/n$ is the spin polarization, and $\epsilon_\mathrm{xc}$ is the exchange-correlation energy per electron in the homogeneous electron gas. We use Tanatar-Ceperley's parameterized form of $\epsilon_\mathrm{xc}$ for the spin-compensated case ($\zeta=0$) and fully spin polarized case ($\zeta=1$).\cite{Tanatar89} $\epsilon_\mathrm{xc}$ for arbitrary spin polarization is obtained by the following interpolation,\cite{Reimann02RMP}
\begin{equation}
\epsilon_\textrm{xc}(n,\zeta) =\epsilon_\textrm{xc}(n,0) +f(\zeta)\left[\epsilon_\textrm{xc}(n,1)-\epsilon_\textrm{xc}(n,0)\right],
\end{equation}
with
\begin{equation}
f(\zeta)=\frac{(1+\zeta)^{3/2}+(1-\zeta)^{3/2}-2}{2^{3/2}-2},
\end{equation}
which is exact for the exchange energy density, but only approximate for the correlation energy density. Recently, Attaccalite, et al.\cite{Attaccalite02} parameterized a new LSDA exchange-correlation functional form based on more accurate quantum Monte Carlo modeling data that includes spin polarization explicitly. The new functional form was shown to be more accurate than Tanatar-Ceperley's in both zero and finite magnetic field when compared to quantum Monte Carlo data.\cite{Saarikoski03} Numerical tests undertaken by us showed, however, that in the regime with which this paper is concerned, the difference between the two forms is minor. In this paper, we will use Tanatar-Ceperley's form. Comparisons with quantum Monte-Carlo calculations for $N \le 12$ and interaction strength parameter $r_s \le 8.0$ have shown that LSDA works well for both the ground state spin and energy in 2D parabolic quantum dots,\cite{Egger99,Pederiva00,Umrigar_private} as well as for many spin excitation energies.\cite{Umrigar_private}

\begin{figure}
\includegraphics[width=2.0in]{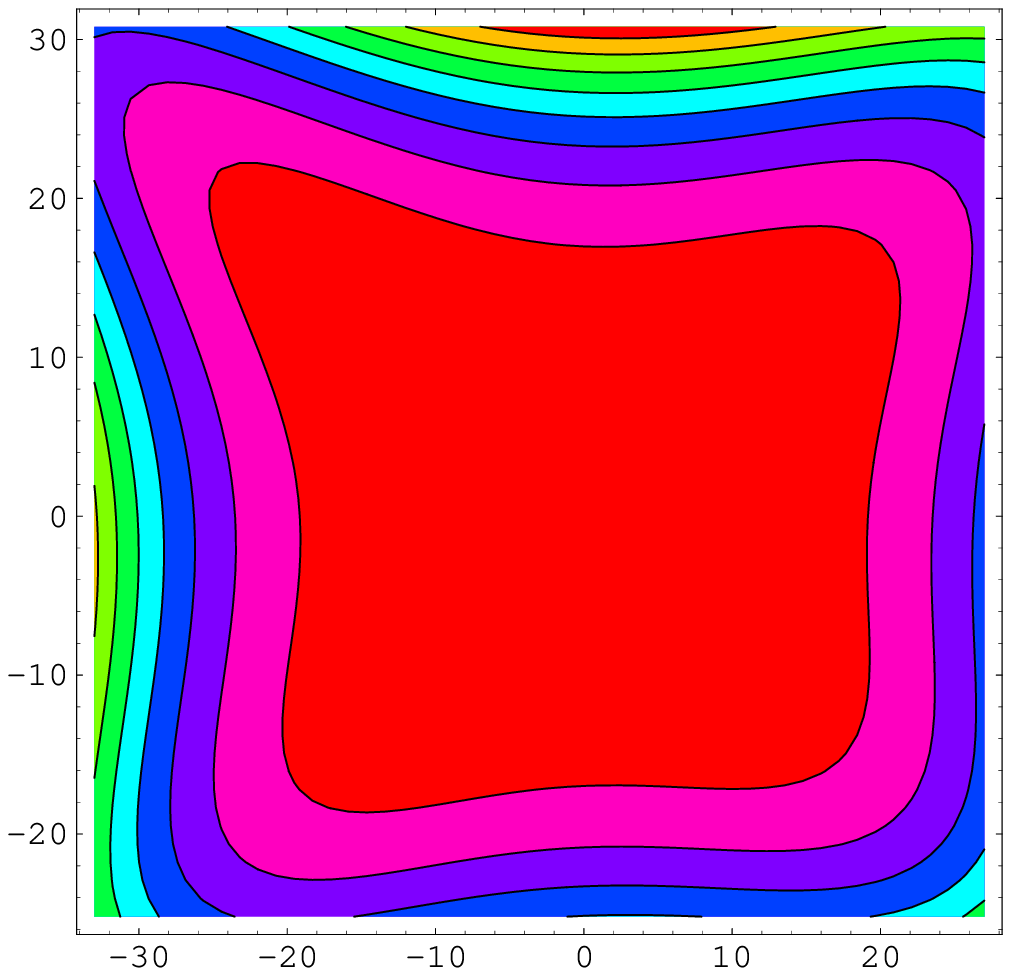}
\includegraphics[width=3.3in]{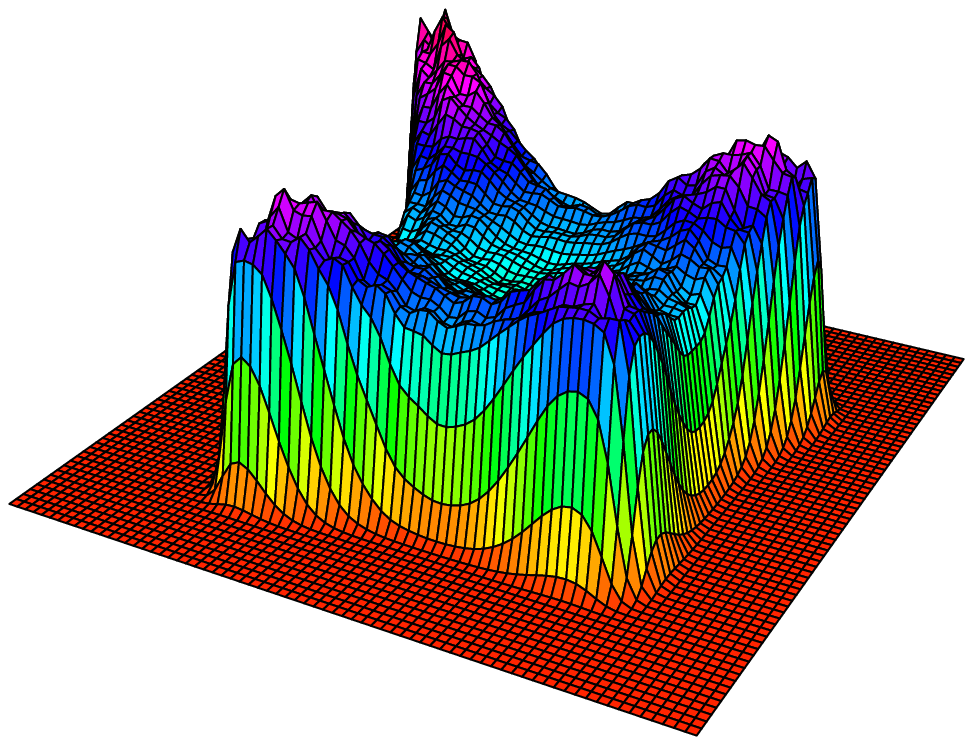}
\caption{\label{fig:qop} (Color online) Upper panel: Contours of $V_\mathrm{ext}$ in QOP system with parameters $a=10^{-4}$, $b=\pi/4$, $\lambda=0.53$, and $\gamma=0.2$. Lower panel: Electron density at $N \!=\! 200$ and $S \!=\! 0$. It is obvious that the electron density in the isolated dot is far from uniform even within the classically allowed area; in  the case shown here, the maximal density is 110\% larger than that in the center.}
\end{figure} 

\begin{figure}
\includegraphics[width=3.2in,clip]{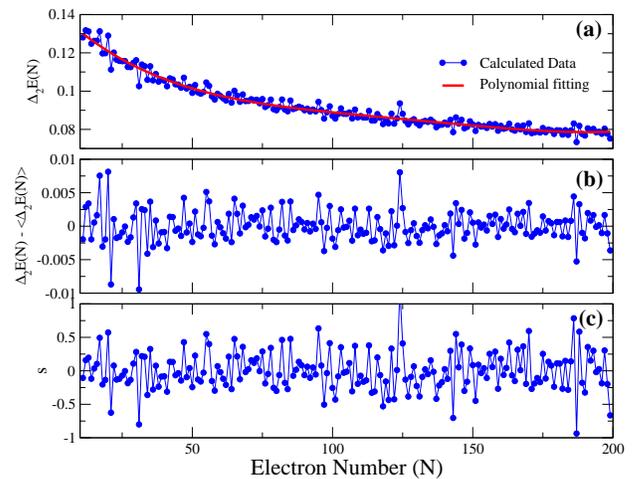}
\caption{\label{fig:qop-data} (Color online) Illustration of data analysis procedure.
(a) The addition energy $ \Delta_2E(N) $ as a function of $ N $ (dotted), and
its $4^{\rm th}$ order polynomial fit that is related to increasing classical
capacitance as the number of electron in the dot increases (solid line).
(b) The fluctuation of addition energy after removing the smooth trend.
(c) Mean-level-spacing-scaled fluctuation of addition energy,
 whose statistical properties are to be studied. The data is from the QOP system
with parameters $a=10^{-4}$, $b=\pi/4$, $\lambda=0.53$, and $\gamma=0.2$
}.
\end{figure}

\subsection{Numerical techniques}

To investigate large electron number regime, we developed efficient techniques for the implementation of KS-SDFT method which have been reported elsewhere.\cite{Jiang03b} Essentially our algorithm includes the following components: (1) We use a particle-in-the-box basis for the representation of KS orbitals and for the action of the kinetic energy operator on the orbital wave functions; (2) The Kohn-Sham total energy is directly minimized by a conjugate gradient method that is based on Teter, Payne and Allan's method\cite{Teter89} but with important modifications, which include a more efficient line minimizaiton scheme and a delayed update of the effective potential;\cite{Jiang03b} (3) We use a Fourier convolution method to calculate Hartree potential efficiently;\cite{Martyna99} (4) A simple one-way multi-grid technique is used to accelerate the convergence.\cite{Lee00} The high efficiency of our method makes the numerical study of statistical properties of large quantum dots with several hundred electrons become computationally feasible.\cite{Jiang03,Jiang03b}

\section{Spin and peak spacing distributions in isolated quantum dots}

We first investigate statistical properties of quantum dots in a
2D isolated model system. Since experimental dots usually have
irregular shapes that result in chaotic classical dynamics, we use
the following quartic oscillator potential (QOP) as the external
confining potential,
\begin{equation}
V_{ext}(\mathbf{r})=a \big\lbrack \frac{x^4}{b}+b  y^4 -2 \lambda x^2
y^2 + \gamma (x^2 y - x y^2)r \big\rbrack.
      \label{eq:qop}
   \end{equation}
Both the classical dynamics and the single-particle quantum mechanics at $\gamma=0$ have been extensively studied:\cite{Bohigas93} the system evolves continuously from integrable to fully chaotic as $\lambda$ changes from 0 to 1.  The parameter $\gamma$ is introduced to break the four-fold symmetry. The prefactor $a$ is used to adjust the flatness of the confining potential and hence the electron density.

\begin{figure*}
\includegraphics[width=6in,clip]{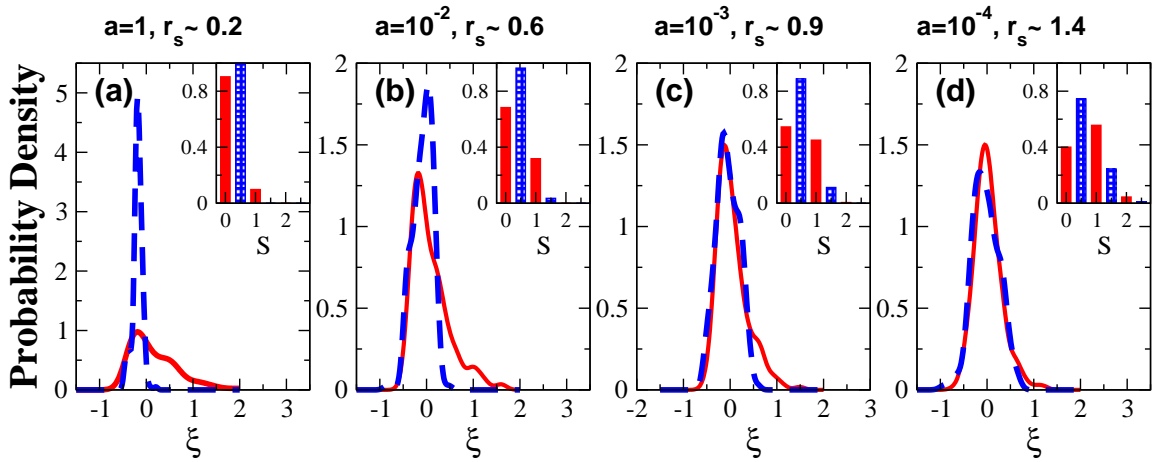}
\caption{\label{fig:qop-distr} (Color online) Distributions of dimensionless peak spacing
[Eq. (\ref{eq:peak-spacing})] for even (solid, red) and odd (dashed, blue) $ N $ in isolated
quantum dots at different interaction strengths. The inset of each figure shows
the corresponding ground state spin distribution for $ N $ even
(solid, red) and odd (shaded, blue). A sliding window is used in estimating
the probability density of peak spacing, yielding a smooth curve
rather than a histogram; the width of the Gaussian window is taken as $
w=\frac{\xi_\mathrm{max}-\xi_\mathrm{min}}{2(1+\mathrm{Log}_2 n)} $
where $ n=200 $ is the number of data points. Note the striking evolution from clearly different even and odd distributions coupled with minimum ground state spin to very similar distributions with a large fraction of high spin ground states.}
\end{figure*}

Fig. \ref{fig:qop} shows the contours of a typical $V_\mathrm{ext}$  used in this study (upper panel), and the corresponding electron density at $N=200$ (lower panel). A noticeable feature is that the electron density in isolated QD systems is not uniform: Electrons form peaks near the boundary region because of classical Coulomb repulsion between electrons.  In real experimental systems, because of the coupling with  the top gate, we expect to see an essentially uniform electron density distribution.  Whether such a difference in electron density will result in significant deviation in statistical properties will be discussed in the next section.

For a given $V_\mathrm{ext}$, the ground state energy $E_\mathrm{gs}$ and spin $S_\mr{gs}$ as a function of $N$ are determined by calculating several spin configurations for each $N$ and selecting the one with minimal energy. The addition energy is then calculated as the second difference of $E_\mathrm{gs}(N)$.  As shown in Fig.~\ref{fig:qop-data}(a), $\Delta_2E(N)$ decreases slowly as $N$ increases. This is mainly a classical effect -- the increasing capacitance as the dot becomes large. Since we are mainly interested in the quantum mechanical effects, the classical smooth trend, denoted $\langle \Delta_2E(N)\rangle$, is fit by a 4-th order polynomial of $N$ and then subtracted from $\Delta_2E(N)$. To compare with experiments, the addition energy is further scaled by the mean level spacing, $\Delta$, which is calculated by $\Delta=2\pi\hbar^2/m^*A_\mathrm{eff}$ with $A_\mathrm{eff}$ estimated as the classical allowed area; the resulting dimensionless spacing is denoted $\xi$,
\begin{equation}
\label{eq:peak-spacing}
\xi(N)\equiv \frac{\Delta_2E(N)-\langle \Delta_2E(N)
\rangle}{\Delta} \,.
\end{equation}
To obtain good statistics, we choose five sets of ($\lambda,\gamma$) -- (0.53, 0.2), (0.565, 0.2), (0.6,0.1), (0.635, 0.15), and (0.67,0.1) -- and for each set of parameters, we calculate a range of electron numbers $N$ from 1 to 200.

Fig. \ref{fig:qop-distr} plots peak spacing distributions in isolated QDs in different interaction regimes; $r_s$ ranges from about $0.2$ to $1.3$, and $N \!=\! 120$-$200$. In the large density or weak interaction regime, $r_s \!\sim\! 0.2$ [Fig. \ref{fig:qop-distr}(a)], the peak spacing distributions for even and odd $N$ are sharply different: the distribution for even $N$ is asymmetric and close to the Wigner surmise while that for odd $N$ is a narrow sharp symmetric peak, which agrees very well with the CI-RMT model. As the interaction strength increases, such an even/odd effect is gradually reduced: the odd $ N $ distribution becomes increasingly wider, and the even $ N $ distribution becomes more symmetric. At $r_s \!\sim\! 0.9$ [Fig. \ref{fig:qop-distr}(c)], the only difference between the two is a weak tail in the even $ N $ distribution at positive spacing. As $r_s$ is further increased to about $1.3$ [Fig. \ref{fig:qop-distr}(d)], which is the usual experimental regime, the even/odd effect is hardly perceptible, and the shape of the distribution is mainly Gaussian-like.

The insets in Fig. \ref{fig:qop-distr} show ground state spin distributions in the corresponding interaction regimes. At $ r_{s} \!\sim\! 0.2 $, the minimal spin configuration dominates in both even and odd $N$ cases, as one expects according to CI-RMT model. But even in this weak interaction regime, the portion of $ S=1 $ for even $ N $ is already significant, $ P_{S=1}=0.1 $. The probability of high-spin ground state increases quickly as the e-e interaction becomes stronger. At $r_s \!\sim\! 1.3$, the high spin ground state is more probable than the minimal spin state.\cite{Jiang03} Such a result is remarkable considering that $ r_{s} \!\sim\! 1.3 $ is only a moderate interaction strength.

\section{Spin and peak spacing distributions in 2D realistic quantum dots}

\subsection{Model}

\begin{figure}
\includegraphics[width=3.2in,clip]{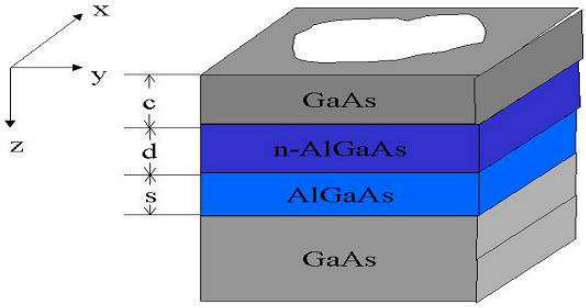}
\caption{\label{fig:rdot} (Color online) Schematic illustration of a realistic quantum dot studied in this paper. The dot is formed based on a GaAs-AlGaAs heterostructure, which consists of (from the bottom) an undoped GaAs substrate, an undoped AlGaAs spacer layer, a n-doped AlGaAs layer, and finally a GaAs cap layer. The widths for the spacer, doped and cap layers are denoted as $s$, $d$ and $c$, respectively. In this paper, $s=15~\mr{nm}$, $d=10~\mr{nm}$ and $c=10~\mr{nm}$ are taken. 
}
\end{figure}

The boundary condition used in isolated quantum dots is very different from that in the experimental systems. Fig. \ref{fig:rdot} illustrates a typical experimental quantum dot structure, which is fabricated based on a GaAs-AlGaAs heterostructure. Numerical modeling of realistic quantum dots\cite{Kumar90,Jovanovic94,Stopa96,Nagaraja97,Matagne02} is still a great challenge in terms of computational efforts. The difficulty mainly comes from the fact that although the volume where electrons are confined is small, the boundary conditions that couple the dot with the external confining gate are defined on a much larger scale. Previous numerical studies of realistic quantum dots considered only small dots with less than a few tens of electrons, except in Ref. \onlinecite{Stopa96} where by reducing Schr\"odinger equations from 3D to multi-component 2D the author was able to study realistically a lateral quantum dot with $N$ up to 100.

\begin{figure}
\includegraphics[width=2.5in,clip]{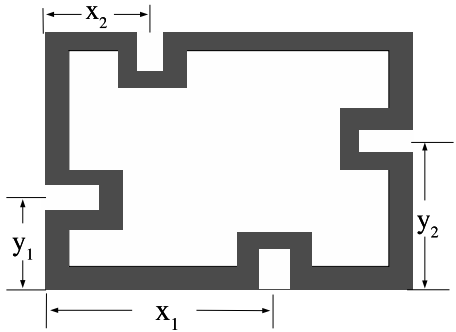}
\caption{\label{fig:rdot-shape} Illustration of the shape of top confining gate
used in this study. A negative voltage is imposed on the shaded region,
depleting the electrons underneath so that the motion of electrons is confined
to a small region. To obtain enough statistics, 20 different irregular
confining shapes are generated by changing $x_1$, $x_2$, $y_1$ and $y_2$
randomly. }
\end{figure}

\begin{figure}
\includegraphics[width=2.0in,clip]{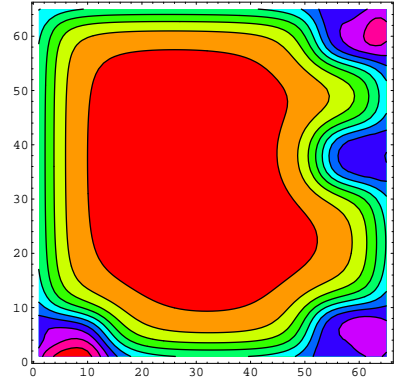}
\includegraphics[width=3.3in,clip]{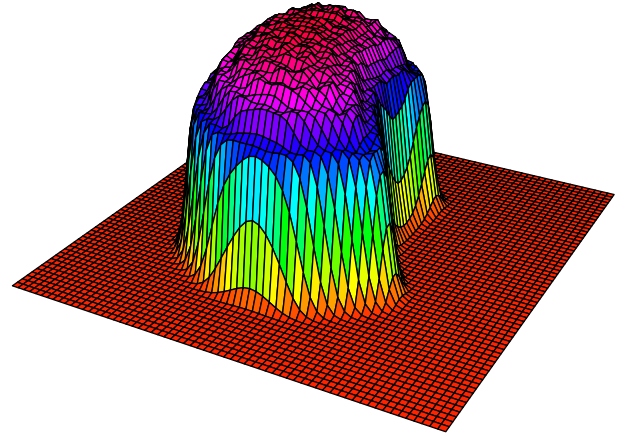}
\caption{\label{fig:rdot-sample} (Color online) Upper panel: Contours of a sample $ V_\mathrm{ext}^\mathrm{R2D} $. With the shape of the confining gate shown in Fig. \ref{fig:rdot-shape}, the confining potential is very irregular; its classical dynamics is believed to be chaotic. Lower panel: Electron density at $N \!=\! 120$. It is much more uniform that that in the isolated QOP shown in Fig. \ref{fig:qop}}
\end{figure}

We note that the most important factor in experimental conditions
that is absent in the isolated quantum dot model is the coupling
between the dot and the top gate. To take that into account, we introduce
semi-infinite boundary conditions.\cite{Davies95,Yakimenko01} Using $ \mathbf{R}=(\mathbf{r},z) $ to denote the 3D spatial
coordinate, $ \mathbf{R}=(\mathbf{r},z) $ where $ z $ is the height in the growth direction
and $ \mathbf{r} $ the coordinates perpendicular to the growth direction, we take
\begin{eqnarray}
\phi(\mathbf{r},z=0) &=&V_g(\mathbf{r})+V_s,\nonumber \\
\frac{\partial \phi}{\partial \mathbf{R}}\big |_{R \to \infty}&=&0,
\end{eqnarray}
where $V_s$ is the Schottky barrier due to surface
states,\cite{Davies88,Davies95} and $V_g(\mathbf{r})$ is the gate voltage imposed on the top surface. In 2D cases, the inclusion of
semi-infinite realistic boundary conditions is quite
straightforward. For $V_\mathrm{ext}$ in Eq. (\ref{eq:KS}),
instead of an analytic form, it is calculated from
\begin{equation}
V_\mathrm{ext}^\mathrm{R2D}(\mathbf{r};z)=\frac{1}{2\pi}\int
\!\! d\mathbf{r}'V_g(\mathbf{r}')
\frac{|z|}{(|\mathbf{r}-\mathbf{r}'|^2+z^2)^{3/2}}+V_\mathrm{QW}(z),
\label{eq:Vconf}
\end{equation}
where $V_\mathrm{QW}(z)$ is the confining potential in
$z$-direction due to the quantum well or heterojunction structure from which
the quantum dot is fabricated. In the case of 2D quantum dots, $z$
is taken as the distance between the top surface and the mean electron height near the middle of the QD, $ z_0 $.
Because of the coupling with the gate, the
Hartree potential has an additional image term,\cite{Davies95,Yakimenko01}
\begin{equation}
V_\mathrm{H}(\mathbf{r})=\int\!
d\mathbf{r}'n(\mathbf{r}') \big[\frac{1}{|\mathbf{r}-\mathbf{r}'|}
-\frac{1}{(|\mathbf{r}-\mathbf{r}'|^2+4z_0^2)^{1/2}} \big] \,.
\end{equation}
This 2D quantum dot model will be denoted R2D to simplify the notation.

Besides the more realistic boundary conditions included in the R2D model, its other advantage is that we can obtain statistics by changing the shape of the confining gate.\cite{Chan95} In this study, we use a confining gate with the shape illustrated in Fig. \ref{fig:rdot-shape}. By taking $x_1$, $x_2$, $y_1$, and $y_2$ randomly, we get a series of irregular confining potentials that are statistically independent as indicated by correlation analysis. Fig.  \ref{fig:rdot-sample} shows the contour of a typical confining potential and a sample electron density for $N=200$. We see that indeed the electron density in R2D is much more uniform than that in the isolated case. The availability of a lot of statistically independent irregular potentials enables us to investigate how spin and peak spacing distributions evolve as a function of $N$ and $r_s$ in a more systematic way.

\begin{figure}
\includegraphics[width=3.37in,clip]{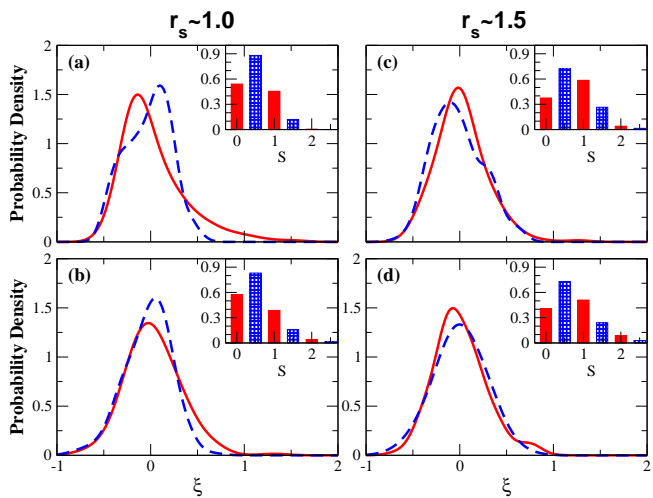}
\caption{\label{fig:qop-r2d} (Color online) Comparison of QOP and R2D models: distributions of conductance peak spacing (solid for even $N$, dashed for odd $N$) and ground state spin (insets) in (a) QOP at $ r_s \!\sim\! 1.0$; (b) R2D at $ r_s \!\sim\! 1.0 $; (c) QOP
at $ r_s \!\sim\! 1.5$; (d) R2D at $ r_s \!\sim\! 1.5 $. For the QOP results $N=80$-$160$, while for the R2D model the additional gate averaging allows a smaller range $N=110$-$130$. Results for the two types of dots are in qualitative agreement.
}
\end{figure}

\subsection{Comparison of distributions from isolated and realistic quantum dots}

Fig. \ref{fig:qop-r2d}  shows spin and peak spacing distributions from isolated and realistic QD systems at two $r_s$ values ($r_s \!\sim\! 1.0$ and $1.5$) within comparable electron number ranges. Qualitatively, they have similar features: At $r_s\!\sim\!1.0$, distributions from both QOP and R2D have a weak tail on the positive side for even $N $ compared to odd $N$, and at $r_s \!\sim\! 1.5$, the tail vanishes. The spin distributions from two types of model systems are also the same within statistical uncertainty. Quantitatively there are indeed some observable differences in both spin and peak spacing distributions, but with the limited statistics available in this study, it is difficult to attribute physical significance to these subtle features. Our study therefore verifies the validity of using model systems to study statistical properties of quantum dots, and provides a direct demonstration of the universal feature of mesoscopic fluctuations.

\begin{figure} \includegraphics[width=3.3in,clip]{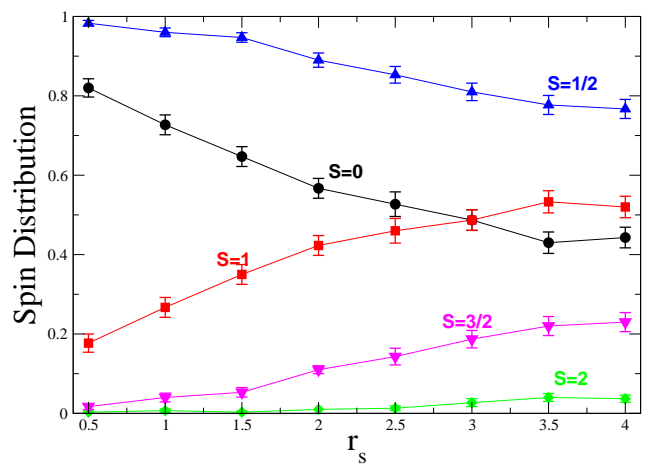} \caption{\label{fig:spin-rs} (Color online) Probabilities of different spin states as a function of $r_s$ for $N\!=\!11$-$40$ in realistic quantum dots (circles: $S\!=\!0$; squares: $S\!=\!1$; diamonds: $S\!=\!2$; up triangles: $S\!=\!1/2$; down triangles: $S\!=\!3/2$). The error bars shown in the plot are calculated from the standard bootstrap method. The ground state spin probabilities saturate at large $r_s$. } \end{figure}

\begin{figure*}
\includegraphics[width=6.5in,clip]{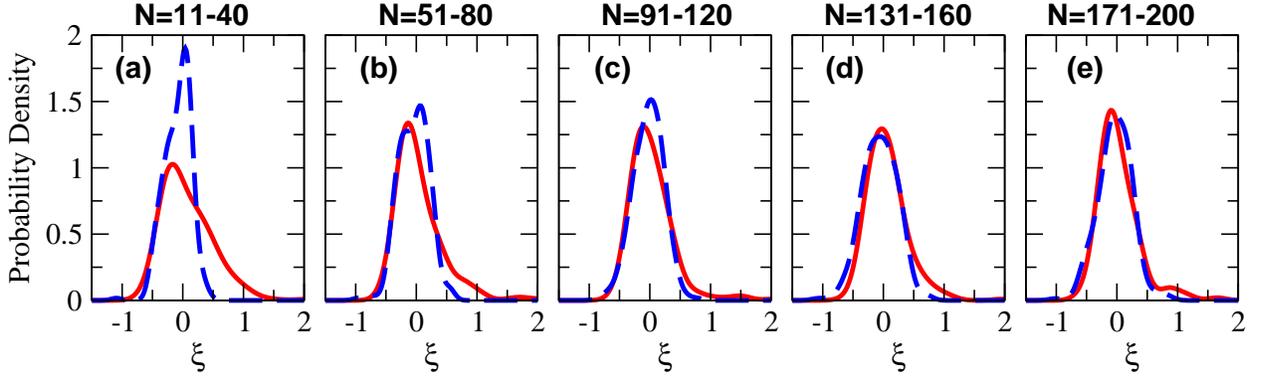}
\caption{\label{fig:EaDistr-Ne} (Color online) Distributions of dimensionless peak spacing for even (solid, red) and odd (dashed, blue) $ N $ in realistic quantum dots in different $N$ ranges at $r_s \!\sim\! 1.0$. There is a clear difference between even and odd distributions for small dots which, however, vanishes for large dots.}
\end{figure*}

\begin{figure}
\includegraphics[width=3.3in,clip]{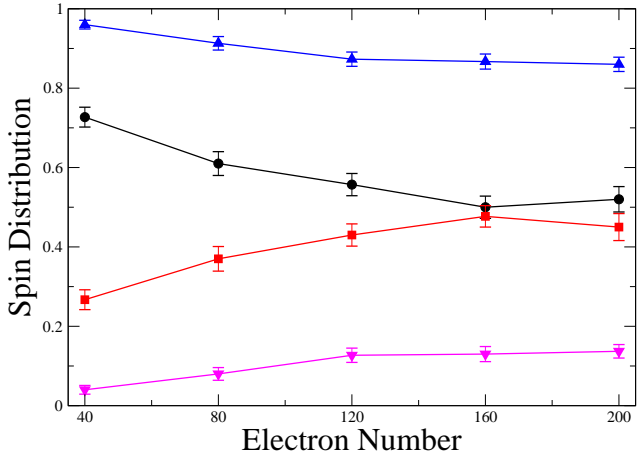}
\caption{\label{fig:Spin-Ne} (Color online) Probabilities of different spin states for different $ N$ ranges at $r_s \!\sim\! 1.0$ (circles: $S\!=\!0$; squares: $S\!=\!1$; up triangles: $S\!=\!1/2$; down triangles: $S\!=\!3/2$). The error bars shown in the plot are calculated from the standard bootstrap method. Note the substantial change from small to large dots.}
\end{figure}

\subsection{Spin distributions as a function $r_s$}

Fig. \ref{fig:spin-rs} shows the spin distribution as a function of the interaction strength in the electron number range $ N\!=\!11$-$40 $. This study extends previous investigations for weak and moderate interaction strength in isolated dots to the stronger interaction regime. In agreement with previous studies, the fraction of high-spin ground states increases as the electron-electron interaction becomes stronger. There are several features that are worth comment. (i)~{\em Crossing:} For even $N$, the probability of $S\!=\!0$ crosses with that of $S\!=\!1$ at $r_s \!\sim\! 3.0 $. (ii)~{\em Saturation:} Probabilities of ground state spins for even $N$ saturate at about $r_s \!\sim\! 3.5$. The spin distribution for odd $N$ is not saturated in the interaction strength range studied here, but the trend indicates that saturation is likely. (iii)~{\em Higher spin states:} The probabilities of $S\!=\!2$ and $S\!=\!5/2$ remain small for all interaction strengths, though $P(S \!=\! 2)$ becomes significant at $r_s>3.0$.

\subsection{Size Effects}
To investigate how statistical properties of quantum dots evolve as a function of electron number $ N $, we calculated peak spacing and spin distributions in different electron number ranges for $N$ up to 200 at $r_s \!\sim\! 1.0$

Fig. \ref{fig:EaDistr-Ne} shows peak spacing distributions in different $N$ ranges. For $N\!=\!11$-$40$, the distributions for even and odd $N$ are obviously different. The difference decreases as $N$ increases, and becomes barely observable for $N\!=\!131$-$160$ and $N\!=\!171$-$200$. Our results show clearly that the conductance peak spacing distribution depends on electron number, and is significantly different  between small ($N<50$) and large ($N>100$) dots.

Fig. \ref{fig:Spin-Ne} shows the trend in ground state spin as $ N $ increases. For this quantity, the dependence on $N$ is even more significant: the probability of high spin ground states increases as $N$ increases. In particular, $P(S\!=\!1)$ increases from 0.27 at $N\!=\!11$-$40$ to 0.45 at $N\!=\!171$-$200$. Another notable feature is that the spin distribution saturates at large $N$.

\subsection{Two-level model analysis}
KS-SDFT calculations reported above reveal some interesting features in conductance peak spacing and ground state spin distribution. To obtain a clearer picture of the underlying physics, and, in particular, to understand the main factors that determine the spin distribution in different regimes, we use a simple two-level model (TLM) to investigate the probability of singlet and triplet ground state spin for even $N$ cases. The same model was used as an interpretive tool in Hirose and Wingreen's\cite{Hirose02} work on spin distributions in small disordered parabolic quantum dots, and in doing that, they attribute saturation of the spin distribution to the effects of off-diagonal interaction matrix elements. Our analysis using single-particle wave functions in truly ballistic chaotic systems,  however, leads to very different conclusions.

The basic idea of the two-level model is to consider a two-electron system with only two
single-particle levels. The interacting matrix elements (IME)
\begin{equation}
V_{ijkl}\equiv \int d\mathbf{r}d\mathbf{r'}\psi_i^*(\mathbf{r}) \psi_j^*(\mathbf{r}')
v_\mathrm{sc}(\mathbf{r},\mathbf{r}') \psi_k(\mathbf{r}') \psi_l(\mathbf{r}).
\end{equation}
are calculated using a screened interaction kernel $v_\mathrm{sc}(\mathbf{r},\mathbf{r}')$ to implicitly account for the effects of all other electrons. In particular, we use the screened interaction obtained from the 2D random-phase approximation (RPA), 
\begin{equation}
v_\mr{sc}(q)=\frac{v_\mr{c}(q)}{1-v_\mr{c}(q)\chi(q)}    
\end{equation}
where $v_\mr{c}(q)=2\pi/q$ is the Fourier transform of the bare Coulomb interaction, and $\chi(q)$ is the susceptibility of the 2D electron gas. It is straightforward to include the screening due to the top gate by using $v_\mr{c}(q)=(2\pi/q)(1-\mr{e}^{-2z_0q})$, but in the regime considered in this paper, the effects of the gate screening on the interacting matrix elements are negligible. 

In this TLM system, one can obtain singlet and triplet state energies either by exact diagonalization (ED) or in the framework of Hartree-Fock (HF) theory (ie., first-order perturbation theory). In TLM-HF, the difference of energy between
singlet and triplet states has the following simple form,
\begin{equation}
\delta E\equiv E_{S=1}-E_{S=0}=\delta\varepsilon-[J_{11}- (J_{12}-K_{12})],\label{eq:TLM-HF}
\end{equation}
where $\delta\varepsilon$ is the single particle level spacing, $J_{ij}\equiv V_{ijji}$ are the direct Coulomb interaction terms, and $K_{12}\equiv V_{1212}$ is the exchange term.

\begin{figure}
\includegraphics[width=3.3in,clip]{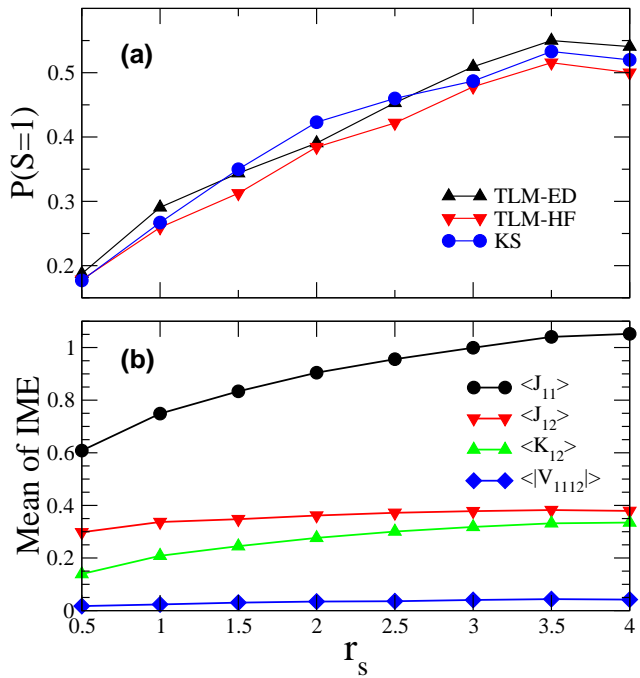}
\caption{\label{fig:TLM-rs} (Color online) Two-level model analysis of spin distribution for $N\!=\!11$-$40$ as a function of interaction strength. (a) Probability of triplet ground state as a function of $r_s$ from TLM-ED (circles), TLM-HF (down triangles) and KS (up triangles). (b) Mean of diagonal interaction matrix elements, $J_{11}$, $J_{12}$, and $K_{12}$, as well as the mean of the off-diagonal matrix element $V_{1121}$, all in units of $\Delta$.}
\end{figure}

\begin{figure}[t]
\includegraphics[width=3.3in,clip]{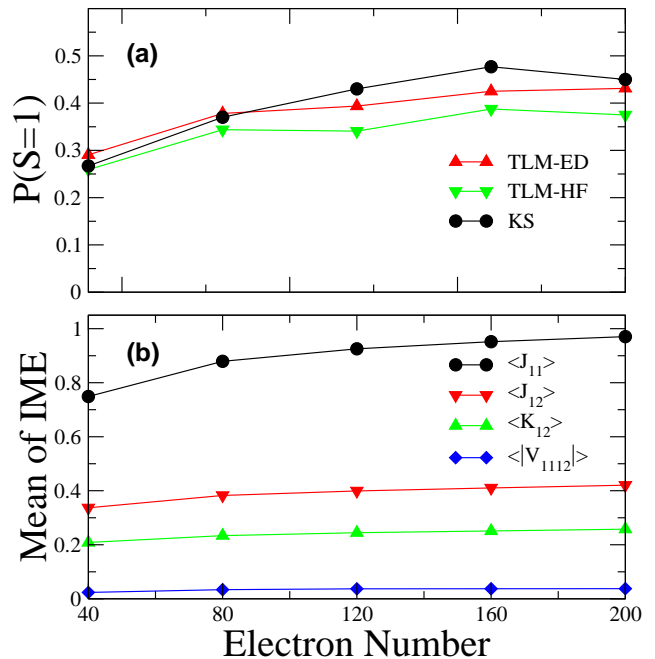}
\caption{\label{fig:TLM-Ne} (Color online) Two-level model (TLM) analysis of spin distribution as a function of $N$. (a) Probability of triplet ground state as a function of $N$ from TLM-ED (circles), TLM-HF (down triangles) and KS (up triangles). (b) Means of diagonal interaction matrix elements, $J_{11}$, $J_{12}$, and $K_{12}$, as well as that of the off-diagonal matrix element $V_{1121}$, all in units of $\Delta$.}
\end{figure}

The main difference between our TLM analysis and Hirose and Wingreen's is in the way to choose single-particle energies and wave functions. In our analysis, we use single-particle eigen-energies and wavefunctions in the extended Thomas-Fermi effective potential\cite{Brack97} that is obtained within the same realistic confining gate structure as in KS-SDFT calculations. Wave functions obtained in this way are guaranteed to be those of ballistic systems, and possible complications related with the disorder strength (ie., localization) can therefore be avoided. We note that the two-level model with this way of calculating interacting matrix elements can be regarded as a simplified version of the model proposed by Ullmo, et al. in Ref. \onlinecite{Ullmo01b}.

Fig. \ref{fig:TLM-rs}(a) shows the probability of finding triplet ground states, $P(S\!=\!1)$, as a function of $r_s$ obtained from TLM-ED, TLM-HF, as well as KS-SDFT calculations. We see a striking agreement between TLM-ED and KS-SDFT results, which is indeed remarkable considering the extreme simplicity of our two-level model compared to KS-SDFT. Both TLM-HF and ED reproduce correctly the saturation of spin distribution at $r_s \ge 3.5$. To explore possible mechanisms for the saturation, Fig. \ref{fig:TLM-rs}(b) plots the mean values of $J_{11}$, $J_{12}$, $K_{12}$, and the off-diagonal term $V_{1121}$, (the other off-diagonal term, $V_{2221}$, is close to $V_{1121}$, and is therefore not plotted here) as a function of $r_s$. It is clearly shown that off-diagonal terms are an order of magnitude smaller than the Coulomb and exchange terms, indicating that the off-diagonal terms are negligible in terms of their contribution to the spin distribution. This is verified by the general agreement between TLM-HF and TLM-ED. On the other hand, Fig. \ref{fig:TLM-rs}(b) shows that though $\langle J_{12} \rangle$ and $\langle K_{12} \rangle$ do saturate at high $r_s$, $\langle J_{11} \rangle$ does not. This indicates that considering mean values of $J_{11}$, $J_{12}$, and $K_{12}$ alone can not explain the saturation in the spin distribution.

We also use the TLM to interpret the effect of finite size on the spin distributions, and the results are plotted in Fig. \ref{fig:TLM-Ne}. The simple two-level model also grasps most of the physics: the probability of the triplet ground state calculated from both TLM-ED and HF increases as $N$ increases, but saturates at large $N$, in agreement with the KS results. We note that the difference between different $N$ ranges is mainly due to wave functions rather than eigenvalues (which are always random-matrix like for our system); for a $N$-electron system, the wave functions involved in the TLM are $\psi_{N/2}$ and $\psi_{N/2+1}$. Our TLM results show clearly that the $N$-dependence of the spin distributions comes mainly from the changing statistics of the wave functions. This is more directly demonstrated in Fig. \ref{fig:TLM-Ne}(b), where mean values of  $J_{11}$, $J_{12}$, $K_{12}$ and $V_{1121}$ are plotted. We note that the $N$-dependence of wave-function statistics is usually neglected in most statistical theories of quantum dots, where the large $N$ limit is usually taken.\cite{Alhassid00RMP,Ullmo01b,Aleiner02,Usaj02} Since many experimental data were obtained with $N$ less than 200, our studies show clearly the necessity of considering finite $N$ effects in interpreting those data.  

\section{Summary}

In this paper, we use Kohn-Sham spin density functional theory to study electron-electron interaction effects in quantum dots. In particular, we calculated conductance peak spacing and ground state spin distributions in an isolated quantum dot with classically chaotic quartic external potential and in a realistic quantum dot with explicit consideration of the coupling between the dot and the external confining gate. In the former case, as an extension of our previous study, we investigated how the statistical properties evolve when the e-e interaction strength increases from the weak ($ r_{s} \!\sim\! 0.2 $) to moderate regimes ($ r_{s} \!\sim\! 1.5 $) for $ N \!=\! 120$-$200 $. In the case of the realistic quantum dots, we first made a direct comparison between isolated and realistic quantum dots, and found that in spite of the difference in the boundary conditions, their statistical properties are qualitatively similar. We further investigated spin and peak spacing distributions in strong interaction regimes with $ r_{s} $ up to 4.0 and their dependence on the electron number $N$. The even/odd effect in peak spacing distributions vanishes as $ r_{s} $ increases and becomes weaker as $ N $ increases.

We close with two comments on our results. First, our study, as well as other theoretical investigations, have made it quite clear that the lack of even/odd pairing in experimental peak spacing distributions is due to the e-e interaction, and the CI-RMT model is therefore not valid for the description of mesoscopic fluctuations in Coulomb blockade peak spacings. In theoretical condensed matter physics, it is well accepted that the e-e interaction at $r_s \!\sim\! 1.5$ can be accurately described by many-body perturbation theory, most notably the RPA model. However, the results obtained from the RPA model combined with a RMT description of single particle properties disagree significantly from those obtained with numerical approaches. Understanding the sources for such discrepancies will deepen our general physical picture of quantum dots. 

Second, our results were obtained under the local spin density approximation within KS-SDFT, whose validity, though well established in 3D bulk material systems through decades of experience, is not necessarily well justified for 2D large quantum dots with irregular confining potential and/or in the strong interaction regime. In the small dots for $N$ up to about 30, it is still computationally feasible to check the validity of LSDA by comparing with more accurate quantum Monte Carlo modeling, but not for the large-dot regime ($N>50$). On the other hand, the statistical quantities investigated above are experimentally accessible. Comparing future experimental data with LSDA predictions will provide a unique chance to check LSDA in real systems.

%%Acknowledgments
\begin{acknowledgments}
We appreciate discussions with G. Usaj and C. J. Umrigar. This work was supported in part by NSF Grant No. DMR-0103003.
\end{acknowledgments}

%\bibliography{nano,rmt,dft,method,remarks}

\begin{thebibliography}{50}
\bibitem[{rem({\natexlab{b}})}]{remark-rs}
\bibinfo{note}{In this paper, we calculated $r_s$ by $r_s \equiv1/\sqrt{\pi
  \bar{n}}$, where $\bar{n}$ is weighted average electron density,
  $\bar{n}=\int n^2(\mathbf{r})d\mathbf{r}/N$. The absolute values of $ r_{s} $
  calculated in this way should not be taken too serious considering that there
  are other ways to calculate $ r_{s} $ in finite systems. For example, in Ref.
  \onlinecite{Hirose02}, $ r_{s} $ is calculated using the density in the
  center point. Experimentally, $ r_{s} $ is estimated from the effective dot
  area.\cite{Patel98, Luscher01}.}
\end{thebibliography}

\end{document}